\documentclass[12pt,letterpaper,reqno]{amsart}
\usepackage{amssymb,amsmath,amsfonts,amsthm,amsbsy,color}
\usepackage[utf8]{inputenc}
\usepackage{natbib}
\usepackage{graphicx}
\usepackage[colorlinks=true,citecolor=blue]{hyperref}%
\usepackage{epsfig,subfigure}
\usepackage{enumerate}
\usepackage{adjustbox}
\usepackage{units}
\usepackage{caption}
\usepackage{setspace,graphicx,setspace,multirow,lscape,longtable,dcolumn}
\usepackage{fullpage}
\usepackage[table, xcdraw]{xcolor}
\usepackage[capposition=top]{floatrow}
\usepackage{booktabs,caption}
\usepackage[flushleft]{threeparttable}
\usepackage{dcolumn}
\usepackage[normalem]{ulem}
\usepackage{arydshln}
\usepackage{xcolor}

\usepackage[color=orange]{changes}

\useunder{\uline}{\ul}{}

\newtheorem*{theorem*}{Theorem}

\graphicspath{ {figures/} }
\title{Predicting Mortality from Credit Reports}

\author{}
\begin{document}
\maketitle

\noindent\textbf{Giacomo De Giorgi}\newline
Geneva School of Economics and Management\newline
University of Geneva\newline
E-mail: \texttt{giacomo.degiorgi@gmail.com}
\vskip 1em
\noindent\textbf{Matthew Harding}\newline
Department of Economics\newline
University of California, Irvine \newline
E-mail: \texttt{harding1@uci.edu}
\vskip 1em
\noindent\textbf{Gabriel F. R. Vasconcelos}\newline
Department of Economics\newline
University of California, Irvine\newline
E-mail: \texttt{gabriel.vasconcelos@uci.edu}\newline
\newline
\noindent
\textbf{Abstract}:
Data on hundreds of variables related to individual consumer finance behavior (such as credit card and loan activity) is routinely collected in many countries and plays an important role in lending decisions. We postulate that the detailed nature of this data may be used to predict outcomes in seemingly unrelated domains such as individual health. We build a series of machine learning models to demonstrate that credit report data can be used to predict individual mortality. Variable groups related to credit cards and various loans, mostly unsecured loans, are shown to carry significant predictive power. Lags of these variables are also significant thus indicating that dynamics also matters. Improved mortality predictions based on consumer finance data can have important economic implications in insurance markets but may also raise privacy concerns.
\newline\newline\newline\newline\newline\newline
\footnotesize{Acknowledgments:   We thank Alexandra Marr who showed interest in an earlier version of this research and contributed preliminary data analyses. De Giorgi thanks Yujung Hwang and Davide Pietrobon for detailed comments, feedback, and assistance along the project, Alex Parret for early feedback, conference participants at the first Workshop on Statistical Learning, and the SNF for  financial support under grant \#100018\_182243.}

\clearpage

\doublespacing

\section{Introduction}

Credit reports have played a central role in consumer lending in the United States since the 1950's and are widely used around the globe. Today credit agencies collect hundreds of variables on each individual who is active in the consumer credit market. The central idea is to capture a consumer's credit worthiness through their past history of borrowing and payments. Most consumers are familiar with the aggregation of this information into a credit score which they often encounter when they apply for a mortgage or a car loan. By capturing the detailed credit history of a consumer, the report provides a unique observational window of the behavior of individuals across a number of core economic activities such as housing, credit card spending, car purchases etc. Credit reports thus capture a multi-modal view of individual behavior. Importantly,  consumers retain some private information concerning their unobservable traits, including their underlying health status, those traits, unobservable  to the researcher,  are however central to the consumer decision process on how to behave in the credit market: i.e. how much to borrow, when, for what purpose, whether to repay, etc. The relevance of such asymmetric information, between parties of a contract, is a centerpiece of contract theory (see \cite{StiglitzWeiss1981}, and \cite{BoltonDewatripont2005}).
At the same time,  various major shocks experienced by the consumer will have both short and long term implications on consumers' behavior which will be indirectly captured in the  credit data.  We use recent developments in machine learning to show that it is possible to predict a seemingly unrelated outcome, mortality, from the data recorded in a consumer report. 
There are three channels through which we believe our approach has a large potential to predict outcomes: i. asymmetric or private information; ii. economic shocks leading the deterioration in health (access to treatment and directly linked to lower resources and the so call death of despair); iii. as a tangible signal of an occurring major health shock which would not be recorded otherwise in easily available data.

We will not try to pinpoint which one is the most likely channel in this project, they could all be at play. We see our work as pioneering the use of large available data in a specific domain to inform decision-making in seemingly unrelated domains.

We also note that the ability to use available Big Data generated in a specific domain to predict outcomes in seemingly unrelated domains   is both an opportunity for innovation and also a potential privacy concern (\cite{harding2014future}).  While not specifically addressing the current COVID-19 crisis, this study highlights the potential of such data to capture the extent to which complex factors such as economic circumstances and choices can be predictive of health outcomes including mortality.

To the best of our knowledge, this is the first paper to study the relationship between credit and mortality directly. However, there is a rich literature that links mortality and health to other economic variables and economic activity. For example, it has been shown that heath is counter-cyclical in the sense that during economic booms health tends to deteriorate \cite{ruhm2000recessions, ruhm2003good, ruhm2005healthy}. Economic growth is associated with increased levels of obesity and a decrease in physical activities, diet quality and leisure time. Also, a reduction of unemployment is associated with a fast increase in coronary heart disease \cite{ruhm2007healthy}. An experimental study showed that higher income is associated with less risk of cardiovascular diseases \cite{wang2019longitudinal}. The relationship however appears to be heterogeneous and it has been documented that recessions are correlated with increases in infant mortality \cite{alexander2011quantifying} and differential across age, race, and education groups \cite{HoynesMillerSchaller2012}.

Research also attempts to identify the underlying mechanisms for the previously mentioned counter-cyclical evidence \cite{stevens2015best}. The evidence is that during economic growth mortality increases mostly amongst elderly women, which suggests that this relation may be linked to factors other than labor. This assumption is supported by the fact that health care may also be counter-cyclical in the sense that staff hiring in nursing facilities increases during recessions. Another theory states that higher mortality is actually related to worst economic conditions during early life \cite{van2006economic}.

Our work contributes this literature of mortality and economics by looking at how credit operations and individual death probabilities are related. We use a very rich data set of microdata from the Experian credit bureau, which is one of the agencies responsible for the credit scores in the United States. We follow a large pool of individuals credit activities through the years and to estimate a modified actuarial life table\footnote{The actuarial life table shows the probability of death within the next year by age and gender. Examples can be found in the Social Security Administration page: www.ssa.gov.} that uses detailed credit variables to get more accurate results. We use machine learning models to deal with the complexity of the data in a pool of more than 2 million individuals (a random 1\% sample of the US population with credit score) and more than a thousand variables in the models. 
To reiterate, our analysis provides an individual level forecasting exercise on mortality making use of available data collected by credit reporting agencies. The ability to predict mortality using routinely performed credit operations opens up to possibility for policy interventions as well as for tailor-made contracts, but at the same time raises privacy concerns over and beyond the classic sharing of sensitive health information

\section{Data}

Our data comes from the Experian Credit Bureau. 
It consists of a yearly sample of more than two million anonymous individuals and 429 variables covering the period between 2004 and 2016. The data is representative for the population of individuals in the US with access to credit.\footnote{Overall over 80\% of the US adult population is represented in the data, with a coverage increasing by age up to age 65 where below 5\% of people do not have a credit score. See for example \cite{LeevanderKlaauw2010}.} The variables are divided in categories such as mortgage loans, car loans, credit cards, installments, etc. The full list of categories is presented in table \ref{tab:groups}. Within the categories there are variables such as number of trades, trade balance, delinquency, etc. A ``deceased flag" in the data will be used to create the mortality variable used in the estimation process. The state of residency is also recorded in the data and will be used separately as discussed below due to potentially confounding effects of geography on mortality given the health disparities in the US. Similar data sets were used to estimate improved credit scores and consumer credit risk using machine learning methods \cite{albanesi2019predicting,khandani2010consumer} but have not been evaluated for their potential to predict outcomes recorded in the data such as mortality.   

The deceased flag is 1 if the individual is dead or died within the reference year. We use this variable to define a mortality variable that is 1 if the individual died in the reference year and 0 otherwise. This is going to be our outcome variable in the machine learning models.

\begin{table}[htb]
\caption{Experian Groups of Variables}
\label{tab:groups}
\begin{tabular}{cc}
\hline
Group & Description                                  \\ \hline
ALJ   & Joint Trades                                 \\
ALL   & ALL Trade Types                              \\
AUA   & Auto Loan or Lease trades                    \\
AUL   & Auto Lease trades                            \\
AUT   & Auto Loan trades                             \\
BCA   & Bankcard Revolving and charge trades         \\
BCC   & Bankcard Revolving trades                    \\
BRC   & Credit Card Trades                           \\
BUS   & Personal Liable Business Loan Line of Credit \\
COL   & Collection trades                            \\
CRU   & Credit Union                                 \\
FIP   & Personal Finance                             \\
ILJ   & Joint Installment Trades                     \\
ILN   & Installment trades                           \\
IQ    & Inquiries                                    \\
MTA   & Mortgage type trades                         \\
MTF   & First mortgage trades                        \\
MTJ   & Joint mortgage trades                        \\
MTS   & second mortgage trades                       \\
PIL   & Personal installment trades                  \\
REC   & Recreational Merchendise trades              \\
REJ   & Joint Revolving trades                       \\
REV   & Revolving trades                             \\
RPM   & Real-estate property managment trades        \\
RTA   & Retail trades                                \\
RTI   & Installment retail trades                    \\
RTR   & Revolving retail trades                      \\
STU   & Student trades                               \\
USE   & Authorized user trades                       \\
UTI   & Utility trades                               \\ \hline
\end{tabular}
\end{table}

\subsection{Estimation Details}

Our estimation uses the full data set available to us consisting of 429 credit variables for 2.2 million individuals covering the period from 2004 to 2016. Besides credit, we also have age and state of residency and we build a variable that counts how many times an individual moved to a different state in the training sample. Our goal is to estimate the probability of death for each individual within the next year using all these variables in the same fashion as the life table published by the Social Security Administration (SSA) conditional only on age and gender. Since our data does not have gender, gender is a protected category and cannot be used in credit decisions and therefore it is not available to us, our benchmark is the probability of death within the next year conditional only on age. Our estimates are not directly comparable with SSA numbers because our sample comprises people with a valid credit score, i.e. they are likely to be wealthier on average than a random sample taken from the entire population, as mentioned earlier for the age groups above 25 the sample covers over 90\% of the US population.

Let $y_{i,t}$ be a binary variable that is 1 if individual $i$ died in period $t$, $X_{i,t}$ contains all credit variables and $Z_{i,t}$ contains variables that are not related to credit, such as age and state of residency. Our estimation framework can be defined by the following equation:

\begin{equation}
\label{eq:1}
    P(y_{i,t+1} = 1 | Z_{i,t}, X_{i,t}, X_{i,t-1},\dots,X_{i,t-k}) =    f(Z_{i,t}, X_{i,t}, X_{i,t-1},\dots,X_{i,t-k}), 
\end{equation}
where $f(\cdot)$ is a general function that will depend on the chosen model and $k$ is the number of lags. Our out-of-sample period is from 2012 to 2016. For each out-of-sample year we estimated several models in the framework of equation \ref{eq:1} using as many lags as possible given the timing of the observation in relation to the length of the available data. The model used to predict mortality in 2012, for example, was estimated with data from 2004 to 2011 (7 lags of the credit variables), which accounts for 3055 variables. The model used to predict mortality in 2016 uses all data from 2004 to 2015, which results in 11 lags and 4771 variables. 

The models we used for the function $f(\cdot)$ are the Random Forest \cite{breiman2001random} and the Gradient Boosting\footnote{Details on the algorithms, implementation and tuning of both models are available in the Supplementary Information.} \cite{friedman2001greedy} in its stochastic version following \cite{friedman2002stochastic} and with early stopping \cite{zhang2005boosting}. The choice of these models is due to several facts. First, they can easily deal with a large number of variables and observations and the algorithms are computationally efficient in very large data sets such as the one we used here, where the outcome variable is an event of very low probability \cite{pike2019bottom}; second, these models are suited for complex data sets where one expects to have many interactions between variables; third, these models are known to be consistent \cite{zhang2005boosting, scornet2015consistency}; finally, both Gradient Boosting and Random Forest are well established machine learning with several successful applications in many different fields \cite{ehrentraut2018detecting, touzani2018gradient, li2018machine}. Our aim is to show the feasibility of the mortality prediction using established techniques which are widely available.

\subsection{Predicting Mortality}

Table \ref{tab:probs} shows the estimated probabilities of dying conditional to the true outcomes from a model trained at $t$ and tested in $t+1$. The values were averaged across all out-of-sample years (2012-2016). The first row is the unconditional probability of death and the remaining rows are the probability of death when the true outcome was death ($y = 1$) and when it was not death ($y = 0$) for the models described in table \ref{tab:abbrev}. The best way to understand this table is to compare for a given model the difference between the probabilities assigned to death when $y=0$ and to survival when $y = 1$. A big gap between these two means that the model is able to allocate higher probabilities to individuals that actually died in year $t+1$.
We notice that conditional on age we have a very small improvement compared to the unconditional model given that the model is assigning bigger probabilities to individuals where the true outcome was death. The inclusion of state of residency does not bring significant improvements however. This is likely due to the fact that while mortality rates differ by geography, in the absence of more detailed location information for the individual, the state of residence by itself carries little predictive power. However, when we evaluate  the models with credit data variables, the improvement is significant, especially for older people. The Gradient Boosting with credit assign twice the probability of death in cases where the individual actually died for most age groups. The Age model, on the other hand assigns probabilities of death less than 10\% larger to people that died in $t+1$ and the models with State dummies do not provide a sizeable improvement. Finally, the Random Forest model performs better, with assigned probability of dying at least 60\% larger for people who actually died at $t+1$. 

We present a more formal comparison between models in table \ref{tab:auc}, which shows the Area Under the Curve (AUC) for all models in all years and the DeLong \cite{delong1988comparing} test to compare their differences. The AUC is the area under the curve of the false positive rate (1-Specificity) against the true positive rate (Sensitivity) for all possible cuts between 0 and 1. If the AUC is larger than 0.5 it means that the model has improved predictive ability compared to the unconditional probabilities. In other words, the AUC tells how much a model is able to distinguish between classes. The values in parentheses in table \ref{tab:auc} show the p-value of the test that compares the Credit Random Forest and the Credit Gradient Boosting to the Age only model. The null hypothesis is that the difference between the AUC of the two tested models is 0. We omitted the models using the state of residence from this table to given that table \ref{tab:probs} already shows that these models do not significantly improve on the models conditional on age alone. The results show that for individuals older than 55 the models with credit data have significantly higher AUCs than traditional actuarial calculations based on age. As we start looking at younger individuals the Random Forest and the Gradient Boosting get closer to the age model to the point where there is no significant difference between them. However, this only happens in general for groups of people younger than 50 is we consider the Gradient Boosting and table \ref{tab:probs} shows that even for younger groups the Gradient Boosting and the Random Forest separate better between the classes. The main explanation to this result is that the number of deaths in the sample for younger groups is very small, as expected, which makes it harder for the models to understand the relation between credit and mortality in a way that can be generalized. Lastly, in table \ref{tab:auc}, the Gradient Boosting algorithm produces slightly higher AUCs than the Random Forest in most cases for older individuals, where the credit models outperform the simpler model based on age. At younger ages, our improved models, while still performing much better than simpler models, suffer from the scarcity of events in the data and somewhat shorter credit histories for younger individuals.

\begin{table}
\caption{Model Abbreviations}
\label{tab:abbrev}
\begin{tabular}{lcc}
\hline
\textbf{Abbreviation} & \textbf{Model}              & \textbf{Variables} \\ \hline
Uncond.              & Unconditional Probabilities &                    \\
Age                  & Conditional Probabilities   & Age                \\
State Lin.           & Linear - Logistic           & Age, State         \\
State GB             & Gradient Boosting           & Age, State         \\
State RF             & Random Forest               & Age, State         \\
Credit GB            & Gradient Boosting           & Age, State, Credit \\
Credit RF            & Random Forest               & Age, State, Credit \\ \hline
\end{tabular}%
\end{table}

\begin{table}
\begin{threeparttable}
\caption{Average Estimated Probability of dying conditional to the true outcomes in $t+1$}
\label{tab:probs}
\resizebox{\textwidth}{!}{%
\begin{tabular}{lcccccccccc}
\hline
                      &     & \multicolumn{9}{c}{\textbf{Age Cohorts}}                                        \\
              &     & 81-100 & 76-80 & 71-75 & 66-70 & 61-65 & 56-60 & 51-55 & 46-50 & 41-45 \\ \hline
Uncond.               &     & 2.99   & 1.52  & 0.98  & 0.61  & 0.41  & 0.25  & 0.16  & 0.10  & 0.07  \\ \hline
Age          & y=1 & 3.27   & 1.54  & 1.00  & 0.63  & 0.42  & 0.25  & 0.16  & 0.10  & 0.07  \\
                      & y=0 & 3.03   & 1.52  & 0.99  & 0.62  & 0.41  & 0.25  & 0.16  & 0.10  & 0.07  \\ \hline
State Lin.      & y=1 & 3.30   & 1.55  & 1.01  & 0.64  & 0.43  & 0.25  & 0.16  & 0.10  & 0.07  \\
                      & y=0 & 3.02   & 1.52  & 0.99  & 0.62  & 0.41  & 0.25  & 0.16  & 0.10  & 0.07  \\ \hline
State GB & y=1 & 3.31   & 1.56  & 1.00  & 0.64  & 0.43  & 0.26  & 0.17  & 0.11  & 0.06  \\
                      & y=0 & 3.00   & 1.52  & 0.99  & 0.62  & 0.41  & 0.25  & 0.16  & 0.10  & 0.07  \\ \hline
State RF   & y=1 & 3.29   & 1.55  & 1.00  & 0.63  & 0.42  & 0.26  & 0.16  & 0.10  & 0.06  \\
                      & y=0 & 3.01   & 1.52  & 0.99  & 0.62  & 0.41  & 0.25  & 0.16  & 0.10  & 0.07  \\ \hline
Credit GB    & y=1 & 4.06   & 0.76  & 0.43  & 0.29  & 0.19  & 0.07  & 0.04  & 0.03  & 0.01  \\
                      & y=0 & 1.93   & 0.42  & 0.25  & 0.15  & 0.09  & 0.04  & 0.02  & 0.01  & 0.01  \\ \hline
Credit RF      & y=1 & 5.56   & 2.68  & 1.82  & 1.28  & 0.87  & 0.51  & 0.34  & 0.26  & 0.15  \\
                      & y=0 & 3.63   & 1.99  & 1.33  & 0.88  & 0.60  & 0.35  & 0.23  & 0.17  & 0.10  \\ \hline
\end{tabular}
}%
\begin{tablenotes}
\small
\item The table shows the estimated probabilities of dying conditional to the true outcomes from a model trained in $t$ and tested in $t+1$. For example, under Prob by Age we have the probability of dying given that the true outcome was death ($y=1$) and the probability of dying given that the true outcome was not death ($y = 0$). The results were averaged across all estimation windows.
\end{tablenotes}
\end{threeparttable}
\end{table}

\begin{landscape}
\begin{table}
\begin{threeparttable}
\caption{Out-of-Sample Area Under the Curve for all Test Years}
\label{tab:auc}
\begin{adjustbox}{max width=\textwidth}
\begin{tabular}{lccccccccccccccccccc}
\hline
{ \textbf{}} & \multicolumn{17}{c}{{\textbf{Out-of-Sample Area Under the Curve}}}                                                                                                                                                         & { \textbf{}} & { \textbf{}} \\
{ }          & \multicolumn{3}{c}{{ Test year = 2012}} & { } & \multicolumn{3}{c}{{ Test year = 2013}} & { } & \multicolumn{3}{c}{{ Test year = 2014}} & { } & \multicolumn{3}{c}{{ Test year = 2015}} & { } & \multicolumn{3}{c}{{ Test year = 2016}} \\
Cohort         & Age         & Credit GB      & Credit RF           &        & Age         & Credit GB      & Credit RF           &        & Age         & Credit GB      & Credit RF           &        & Age         & Credit GB      &Credit RF           &        & Age    &Credit GB        & Credit RF              \\ \cline{1-4} \cline{6-8} \cline{10-12} \cline{14-16} \cline{18-20} 
81-100          & 0.587       & 0.659         & 0.657        &        & 0.576       & 0.673         & 0.678        &        & 0.578       & 0.693         & 0.684        &        & 0.576       & 0.682         & 0.663        &        & 0.562  & 0.703           & 0.683           \\
                &             & (0.000)       & (0.000)      &        &             & (0.000)       & (0.000)      &        &             & (0.000)       & (0.000)      &        &             & (0.000)       & (0.000)      &        &        & (0.000)         & (0.000)         \\
76-80           & 0.523       & 0.604         & 0.619        &        & 0.525       & 0.607         & 0.607        &        & 0.526       & 0.608         & 0.610        &        & 0.542       & 0.608         & 0.603        &        & 0.542  & 0.612           & 0.609           \\
                &             & (0.000)       & (0.000)      &        &             & (0.000)       & (0.000)      &        &             & (0.000)       & (0.000)      &        &             & (0.000)       & (0.000)      &        &        & (0.000)         & (0.000)         \\
71-75           & 0.518       & 0.596         & 0.620        &        & 0.521       & 0.595         & 0.619        &        & 0.522       & 0.592         & 0.615        &        & 0.534       & 0.606         & 0.595        &        & 0.546  & 0.587           & 0.594           \\
                &             & (0.000)       & (0.000)      &        &             & (0.000)       & (0.000)      &        &             & (0.000)       & (0.000)      &        &             & (0.000)       & (0.000)      &        &        & (0.000)         & (0.000)         \\
66-70           & 0.545       & 0.603         & 0.603        &        & 0.543       & 0.619         & 0.611        &        & 0.528       & 0.612         & 0.592        &        & 0.531       & 0.612         & 0.607        &        & 0.535  & 0.617           & 0.617           \\
                &             & (0.000)       & (0.000)      &        &             & (0.000)       & (0.000)      &        &             & (0.000)       & (0.000)      &        &             & (0.000)       & (0.000)      &        &        & (0.000)         & (0.000)         \\
61-65           & 0.491       & 0.583         & 0.592        &        & 0.519       & 0.598         & 0.594        &        & 0.531       & 0.602         & 0.604        &        & 0.546       & 0.594         & 0.605        &        & 0.528  & 0.599           & 0.592           \\
                &             & (0.000)       & (0.000)      &        &             & (0.000)       & (0.000)      &        &             & (0.000)       & (0.000)      &        &             & (0.000)       & (0.000)      &        &        & (0.000)         & (0.000)         \\
56-60           & 0.511       & 0.619         & 0.568        &        & 0.550       & 0.610         & 0.574        &        & 0.539       & 0.595         & 0.595        &        & 0.545       & 0.591         & 0.582        &        & 0.537  & 0.594           & 0.590           \\
                &             & (0.000)       & (0.001)      &        &             & (0.000)       & (0.152)      &        &             & (0.000)       & (0.001)      &        &             & (0.002)       & (0.015)      &        &        & (0.000)         & (0.000)         \\
51-55           & 0.530       & 0.616         & 0.601        &        & 0.502       & 0.566         & 0.580        &        & 0.545       & 0.591         & 0.579        &        & 0.527       & 0.591         & 0.573        &        & 0.541  & 0.578           & 0.562           \\
                &             & (0.000)       & (0.001)      &        &             & (0.002)       & (0.000)      &        &             & (0.018)       & (0.083)      &        &             & (0.000)       & (0.013)      &        &        & (0.040)         & (0.228)         \\
46-50           & 0.559       & 0.650         & 0.595        &        & 0.515       & 0.617         & 0.558        &        & 0.545       & 0.599         & 0.581        &        & 0.553       & 0.603         & 0.583        &        & 0.510  & 0.610           & 0.577           \\
                &             & (0.001)       & (0.184)      &        &             & (0.000)       & (0.085)      &        &             & (0.042)       & (0.148)      &        &             & (0.028)       & (0.216)      &        &        & (0.000)         & (0.003)         \\
41-45           & 0.539       & 0.557         & 0.551        &        & 0.536       & 0.574         & 0.597        &        & 0.546       & 0.594         & 0.537        &        & 0.549       & 0.549         & 0.527        &        & 0.536  & 0.592           & 0.575           \\
                &             & (0.582)       & (0.706)      &        &             & (0.213)       & (0.043)      &        &             & (0.108)       & (0.746)      &        &             & (0.973)       & (0.438)      &        &        & (0.083)         & (0.187)         \\ \hline
\end{tabular}
\end{adjustbox}
\begin{tablenotes}
\small
\item The table shows the out-of-sample Area Under the Curve (AUC) for all test years, cohorts and models. Values in parenthesis are \\ p-values from \citet{delong1988comparing} test for the Random Forest and the Gradient Boosting against the probabilities conditional on \\ age only. The null hypothesis is that the difference between the AUC of both models is 0.
\end{tablenotes}
\end{threeparttable}
\end{table}

\end{landscape}

\begin{figure}[htb]
\caption{Out-of-Sample Area Under The Curve - 2012}
\label{f:auc2012}
\includegraphics[width=0.95\textwidth]{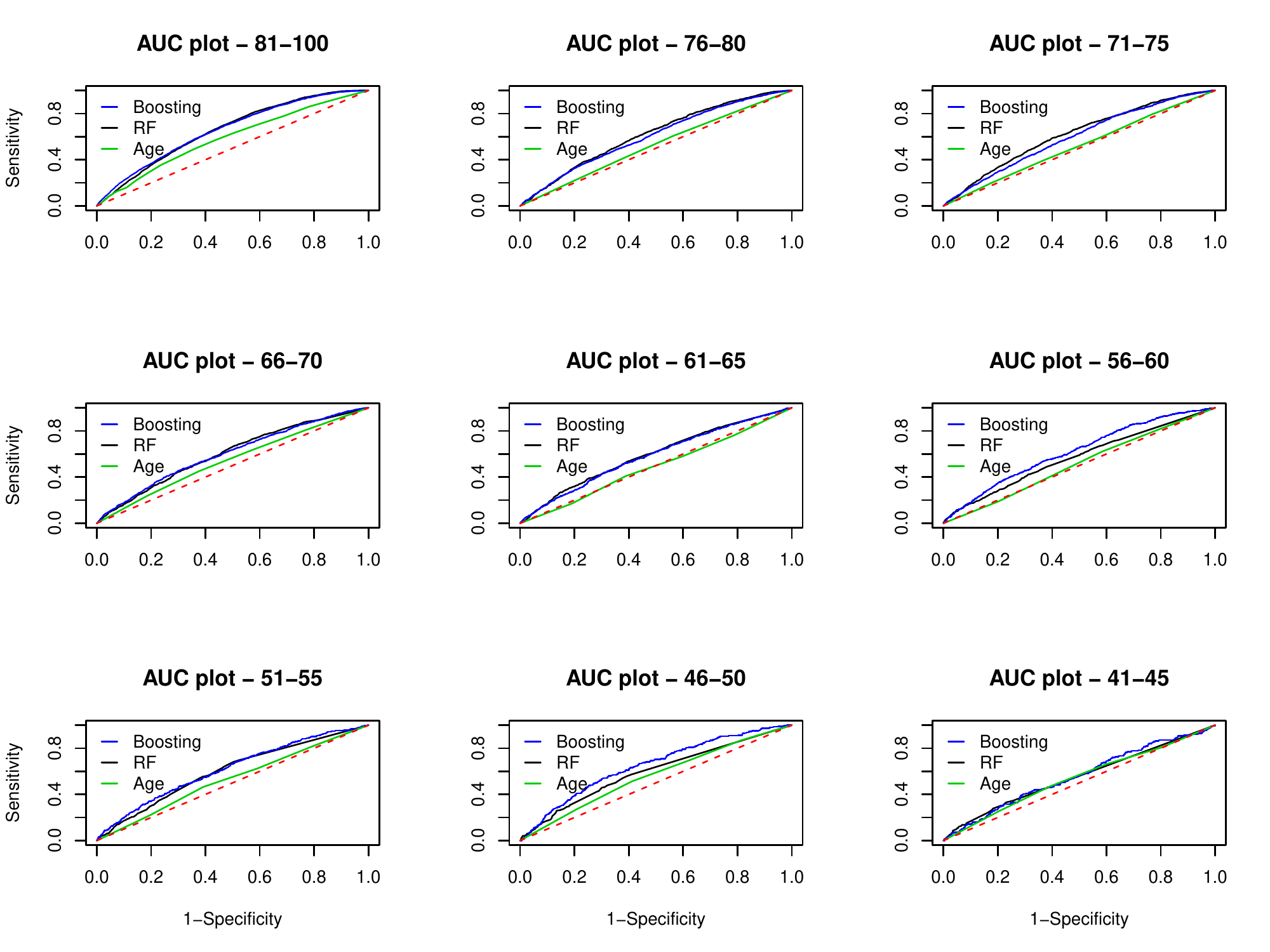}
\end{figure}

\begin{figure}[htb]
\caption{Out-of-Sample Area Under The Curve - 2016}
\label{f:auc2016}
\includegraphics[width=0.95\textwidth]{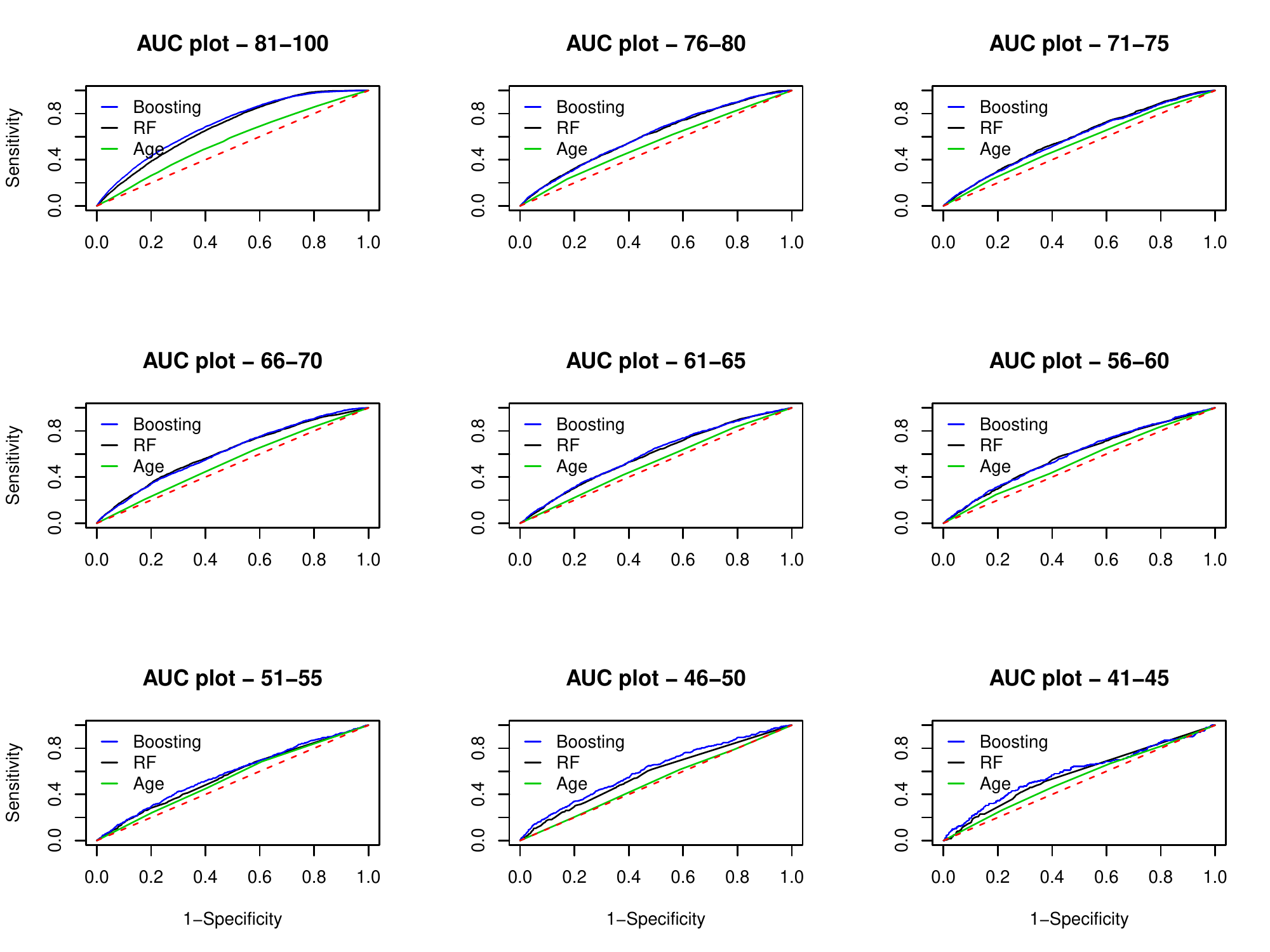}
\end{figure}

The same results presented in table \ref{tab:auc} can be observed in more details in figures (\ref{f:auc2012} and \ref{f:auc2016}). These figures show the AUC plots for years 2012 and 2016 and all age cohorts. It is very clear that all curves deviate more from the $45^{\circ}$ line for cohorts with older people. The small deviations in younger people are likely due to the low mortality rate for these cohorts resulting in a number of deaths in each year in the sample that is insufficient to successfully apply these algorithms. The same behavior persists through all out-of-sample years. The number of lags does not seem to increase the performance of the models. The 2012 model had 7 lags and the 2016 model had 11 lags. 
We go back to the discussion on predictive value of distant lags in the next section.

\subsection{Variable Importance}

\begin{figure}[htb]
\caption{Variable Importance by Groups and Age Cohorts}
\label{f:importance_group}
\includegraphics[width=0.95\textwidth]{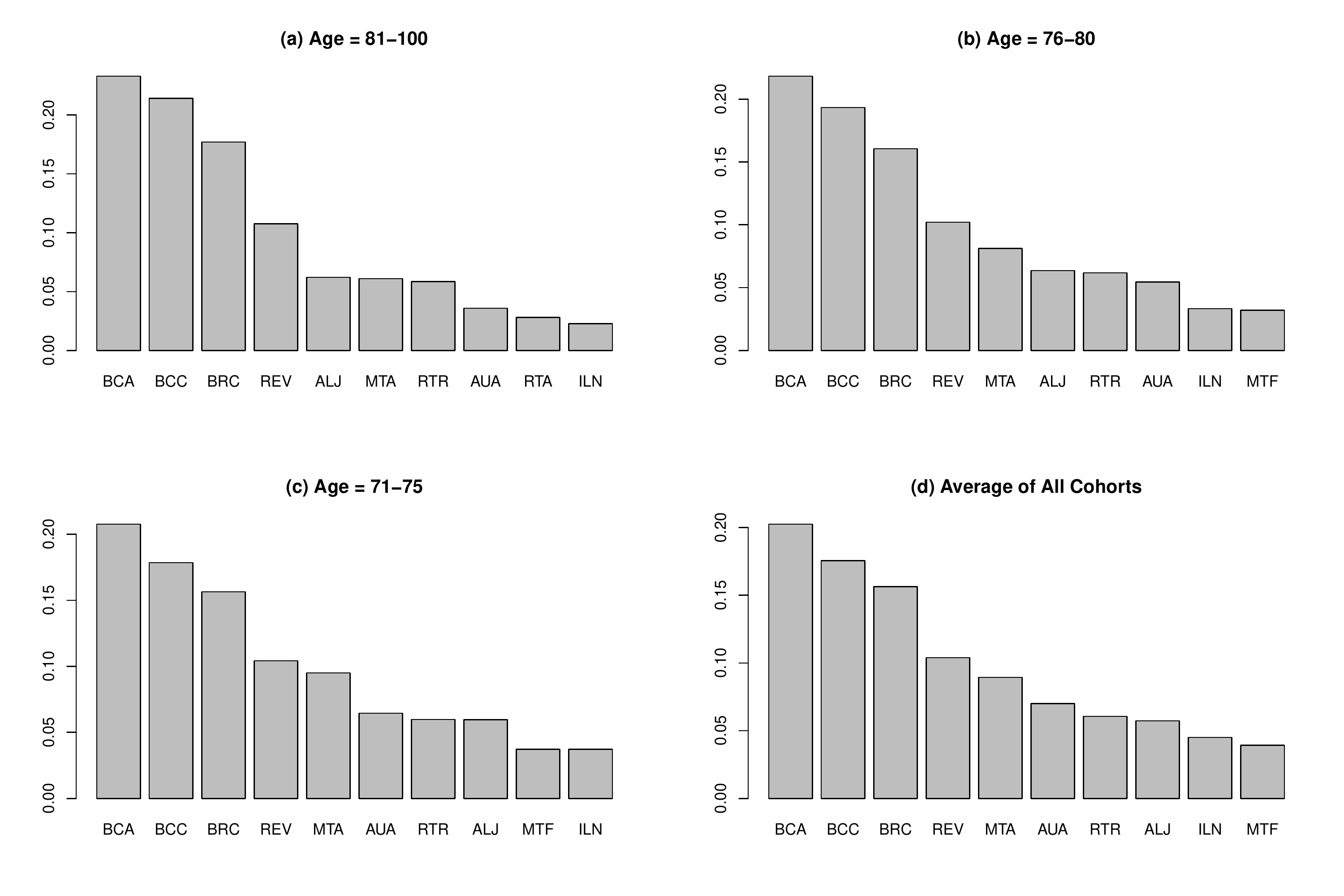}\\
\floatfoot{The figure shows the relative importance of the 10 most relevant groups of variables based on the Experian classification for several age cohorts. Values were adjusted so that the sum of all bars is equal 1. The importance was measured as the contribution of each variable to reduce the model prediction error.}
\end{figure}

In this subsection we address which variables are more relevant to reduce the prediction errors in our models. We present results for the Gradient Boosting, which was slightly more accurate in general than the Random Forest. Figure \ref{f:importance_group} shows the relative importance between the 10 most relevant groups of variables by age cohorts. The list of all groups is presented in table \ref{tab:groups}. The four most important groups (BCA, BCC, BRC and REV) are related to credit and bank cards and revolving trades, unsecured credit. Mortgage and joint trades variables (MTA, MTF, ALJ) are also of some importance. Auto loan (AUA) variables also appear in the most important groups for all cohorts. As for the difference between age cohorts, mortgage and auto related loans are less relevant for older people than for younger ones. Of less importance but still worth noting are Installment trades. These results seem intuitive given that credit cards and other revolving trades are classes of credit that can change very fast with individual circumstances. Furthermore, mortgage decisions are very central to many younger individuals and are driven by long run expectations. Likewise, one would expect mortgage and auto loans to be less important for older individuals.  Overall, though the pattern of variable importance is remarkably consistent across age cohorts. The only exception appears to be the cohort of consumers over 80 years old. It is possible that many of these consumers ``simplify" their financial life. For example mortgages are paid off, the elderly tend to reduce consumption in many areas and thus need to rely less on credit cards to smooth out consumption over time etc.

\begin{figure}[htb]
\caption{Variable Importance by Groups and Lags}
\label{f:importance_lags}
\includegraphics[width=0.95\textwidth]{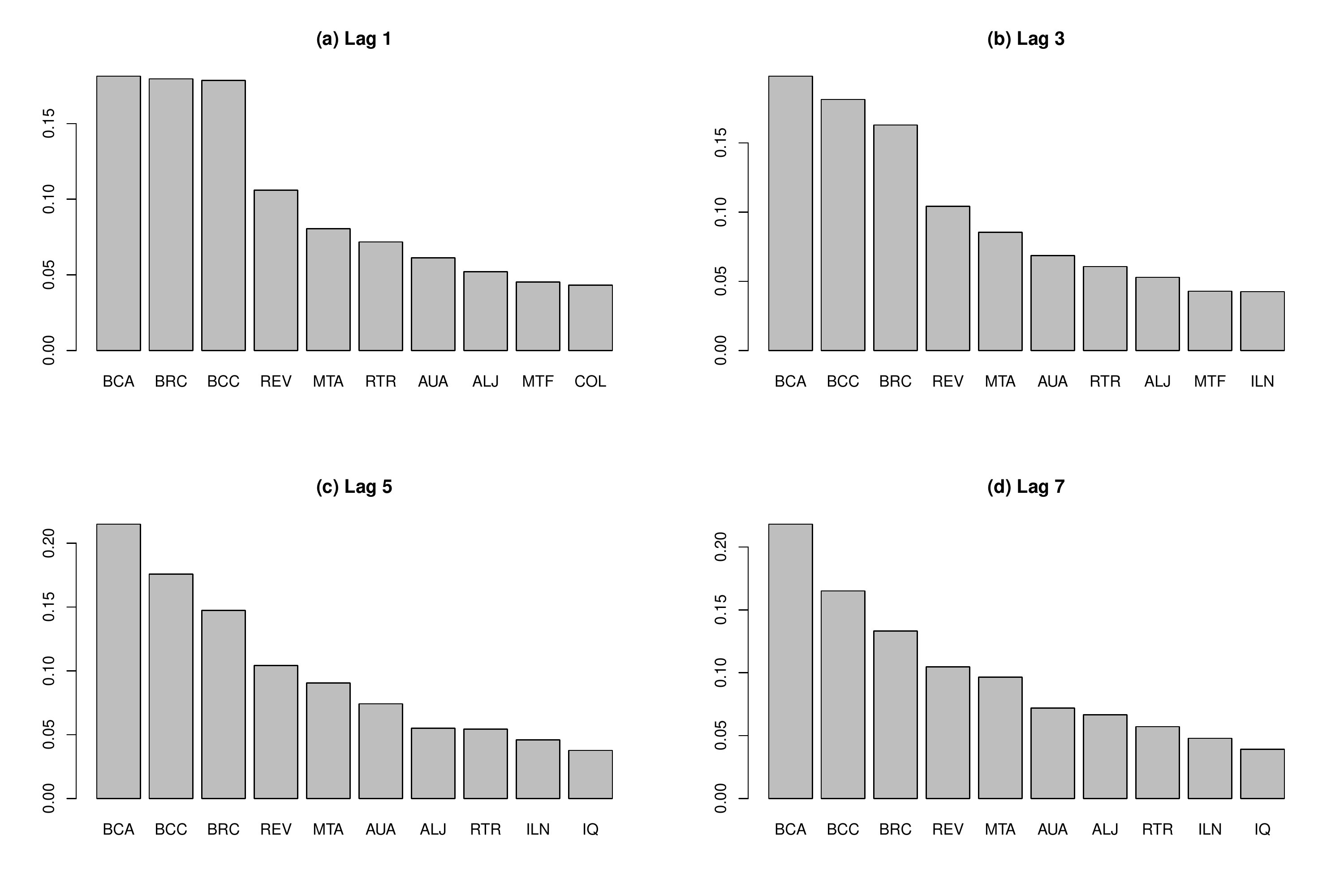}\\
\floatfoot{The figure shows the relative importance of the 10 most relevant groups of variables based on the Experian classification for lags 1-7. Values were adjusted so that the sum of all bars is equal 1. The importance was measured as the contribution of each variable to reduce the model prediction error.}
\end{figure}

Figure \ref{f:importance_lags} shows the relative importance between the 10 most relevant groups of variables by lags. What is quite interesting is that for the class of variables that are the most responsive in the short run, they are also very central for the long lags. For example, the 5-years lags overall appears to have about a 1/3 importance with respect to the 1-year lag, yet that rules out some simple story of running into default because of death or health shocks only. It seems that part of the predictive power, especially in long-lags, is due to individual revelation of their types rather than reaction to shocks. This is consistent with the hypothesized mechanism where credit behavior reveals the particular individual type, a story of private information rather than economic shocks. 

\begin{figure}[htb]
\caption{Importance by lags}
\label{f:importance_lag_only}
\includegraphics[width=0.5\textwidth]{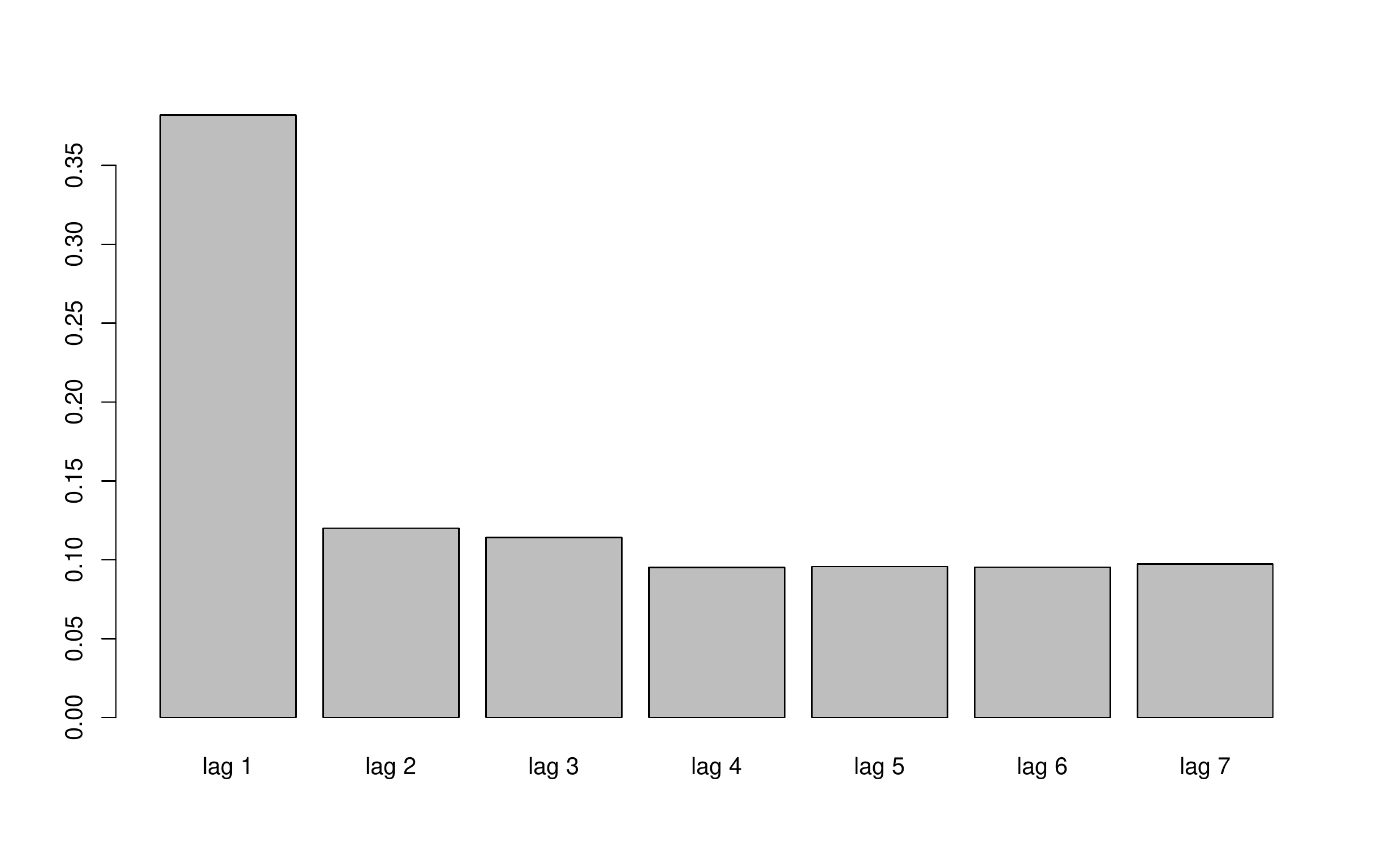}\\
\floatfoot{The figure shows the relative importance between lags 1 to 7 averaged across all models. Values were adjusted so that the sum of all bars is equal 1. The importance was measured as the contribution of each variable to reduce the model prediction error.}
\end{figure}

Note that the most relevant variables are the same as the ones in figure \ref{f:importance_group}. However, we can see a bigger change in their relative importance as we move from lag 1 to lag 7. Inquiries (IQ) appear only for bigger lags (5 and 7) and in lag 1 we have collection trades (COL) as a relevant group. Finally, figure \ref{f:importance_lag_only} shows the relative importance between lags. Lag 1 clearly dominates the remaining lags but the overall importance of lags 2 to 7 combined is bigger than lag 1 alone. This is evidence that the relationship between credit and mortality has a long term component and financial positions taken several years in the past may be predictive of mortality (and potentially other health outcomes) in the present.  We take this as suggestive of sizeable private information retained by consumers on their health status, i.e. the predictive power of distant past behavior could be reflecting private information on one's general health and life-expectancy rather than short term shocks. It is less likely that these results are consistent with a simple story of a health shock leading to bankruptcy or an income shock reducing the individual ability to access health care. It is interesting to note that while there is a sharp reduction in the importance of the variables from lag 1 to lag 2, the importance after does not decline perhaps as rapidly as we might expect. We think that this is consistent with a process where credit reports capture both immediate shocks and more persistent effects. For example mortality is driven both by sudden health events such as a stroke and long run uncontrolled blood pressure which is related to lifestyle choices such as a sedentary life and bad nutrition.

\section{Conclusion}

We have shown that data routinely collected by credit bureaus such as Experian in the US have substantial power in predicting mortality at the individual level. We employed data on over 2 million individuals and 429 credit related variables to estimate Gradient Boosting and Random Forest models for the probability of an individual dying within the next year. Our models significantly outperform actuarial tables conditional on age in terms of AUC, which shows the model ability to distinguish between classes. Moreover, the Gradient Boosting assigns probabilities of death twice as big to individuals that actually died at $t+1$ against an increase of less than 10\% in the actuarial age model. A limitation of our study is that we do not observe gender in our data which is commonly reported in actuarial tables, but it is unavailable in credit data. The predictive performance of our algorithms improves with the age of the individual. It is not clear whether this is due to the increased information content of credit data for older Americans or whether this is an artefact of the estimation strategy and the relatively low number of deaths in younger cohorts. The measured variable importance seems to suggest that the results are driven by the measured changes in credit and bank cards (such as balances and payment amounts). Additional variable groups related to mortgage activity and other loans are also predictive but to a lesser extent.

The current study is not meant to fully uncover the underlying mechanism which are likely to be complex and potentially subject to many feedback cycles between credit behavior and health. Our insights are however consistent with the current state of knowledge which documents correlations between economic shocks and health outcomes, and potentially with individuals retaining a substantial amounts of private information on their health status. These correlations appear to be highly predictive of mortality outcomes. It is important to note that lags of the credit variables are also predictive which is indicative of both a short and long run component of the relationships between health and consumer finance. 

The documented predictive power of credit variables for individual mortality has a number of implications. From an economic perspective, mortality predictions play an important role in a number of markets such those for life insurance and reverse mortgages. Life expectancy calculations are also key in legal proceedings which rely on evaluations of the expected value of life. Our study shows that actuarial tables that are usually relied upon can be significantly improved upon at the individual level using relatively common if proprietary data collected routinely for most Americans.  Thus even without access to any sort of health information on the individual, health outcomes such as mortality can in fact be inferred from credit data.

\clearpage

\bibliographystyle{agsm}

\bibliography{mortality.bib} 

\clearpage

\appendix

\section{ }

This supplementary text contains details on the Machine Learning models used in the paper, the benchmark models and some additionally results. It is divided in to four sections. The first section presents the benchmark models, the second section shows the Machine Learning (ML) models, the third section discuss some technical details of the models and the final section shows some additional results and statistics. 

\subsection{Benchmark Models}

We have three benchmark models that do not use credit data:

\begin{itemize}
    \item The unconditional model uses only the number o individuals that died in the year $t$ to estimate the probability of dying in year $t+1$.
    \item The Age model does the same as the unconditional model. However, probabilities are calculated conditional on the age. 
    \item The linear logistic model uses a logistic regression with age and state of residency to obtain the probabilities. 
\end{itemize}

\subsection{Machine Learning Models}

The Machine Learning models used in this research are Gradient Boosting \cite{friedman2001greedy} and Random Forest \cite{breiman2001random}. Both models are based on Classification and Regression Trees (CART) \cite{breiman2017classification}. CART is a nonparametric model that approximates a nonlinear function with local predictions using recursive partitioning of the space of the predictor variables. For example, if we are modeling the mortality using age and State of residency one partition could be $age > 60$ and a second partition could be $State = CA$. In this case, people older than 60 years old that live in California would be classified with the same mortality probability.  

Random Forests (RF) \cite{breiman2001random} use CART models in a Bootstrap Aggregating (Bagging) \cite{breiman1996bagging} algorithm with some adjustments in the regression trees. It consists on the estimation of a large number of regression trees on bootstrap samples where the final result will be the average of all trees in a regression problem or each tree will count as a vote in a classification problem. Additionally, when searching for a new partition variable the Random Forest looks only at a random subset of all candidate variables in order to introduce more randomness and to increase the difference between individual trees. 

Gradient Boosting (GB) \cite{friedman2001greedy} in the context that we used also relies on CART as a building block. However, instead of independent trees like in the Random Forest, GB algorithms estimate small trees on the pseudo-residuals of the current iteration of the model. Suppose we are on iteration $m$. First, we calculate the residuals $u_{i,m} = y_i - \phi_{i,m-1}$. Then we estimate a CART model on $u_{i,m}$ to obtain the next step $\phi_{i,m} = \phi_{i,m-1} + v \rho_{m} \hat{u}_{i,m}$, where $\rho_m$ is the step size estimated within the algorithm and $v$ is the learning rate chosen by the user. This procedure is repeated for $M$ iterations.

\subsection{Tuning of ML Models}

For the Random Forest we estimated models with ranger \cite{rangerpk} package in R. We used 500 trees with minimum terminal node size of 5 observations to limit the tree size. The number of variables tested in each new partition was 1/3 of all potential variables. 

The Gradient Boosting was estimated with the xgboost \cite{xgb} package in R. The Gradient Boosting requires more parameter tuning than the Random Forest. We used 500 iterations in the estimation, the objective functions was the logistic function, the subsample proportion for the stochastic estimation was 0.25 and every time a new node was created we looked only at 1/3 of the variables randomly selected in each split. The learning rate was set to 0.2. All the remaining parameters were used as the package default. The subsampling and the random selection of variables are used to reduce over-fitting and the learning rate of 0.2 is small enough to avoid over-fitting and big enough to reach convergence in 500 iterations. We tested different setups and the results were similar.

\subsection{Other Results and Robustness}

\begin{figure}[htb]
\caption{Out-of-Sample Area Under The Curve - 2013}
\label{f:auc2013}
\includegraphics[width=0.95\textwidth]{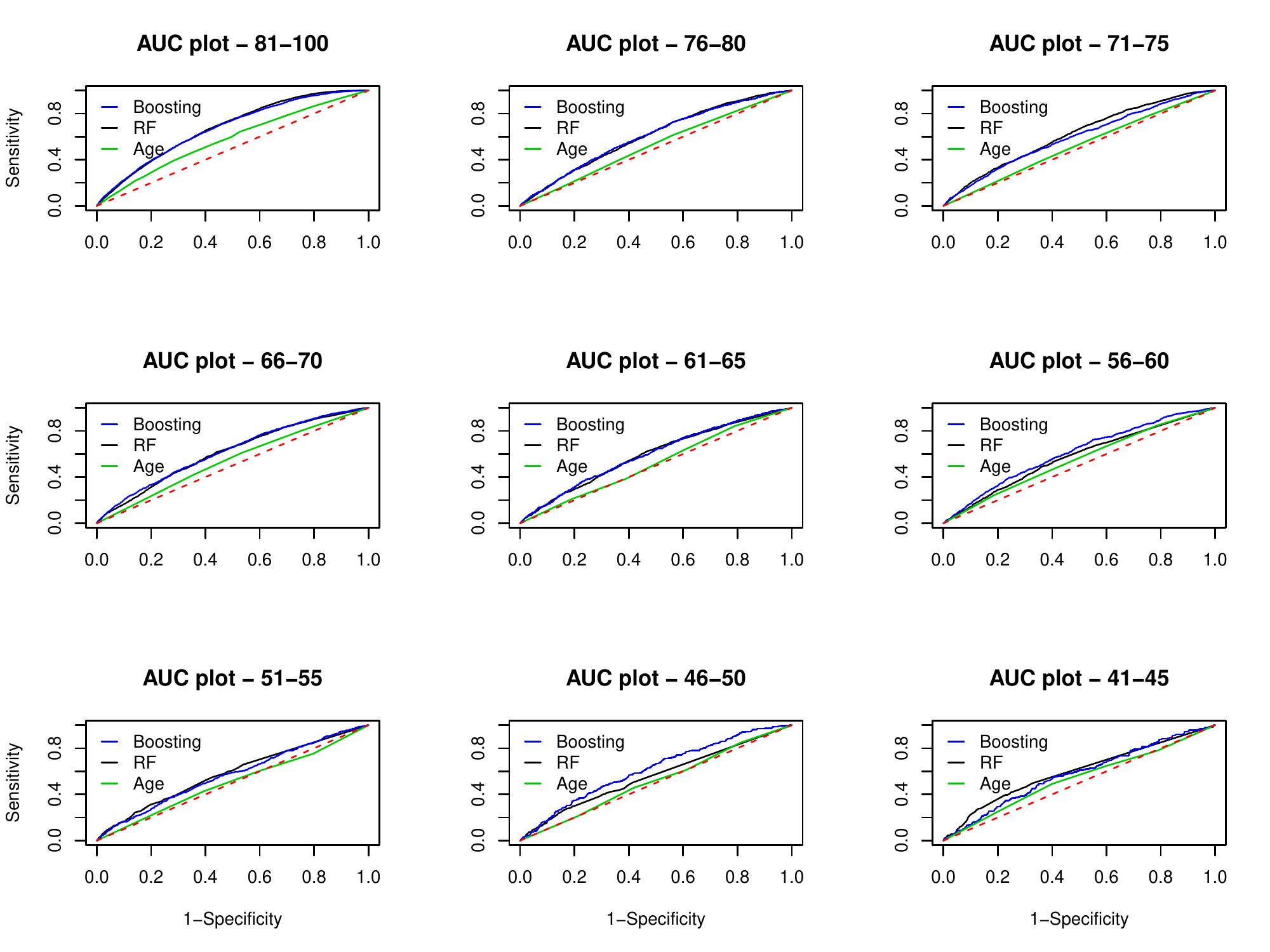}
\end{figure}

\begin{figure}[htb]
\caption{Out-of-Sample Area Under The Curve - 2014}
\label{f:auc2014}
\includegraphics[width=0.95\textwidth]{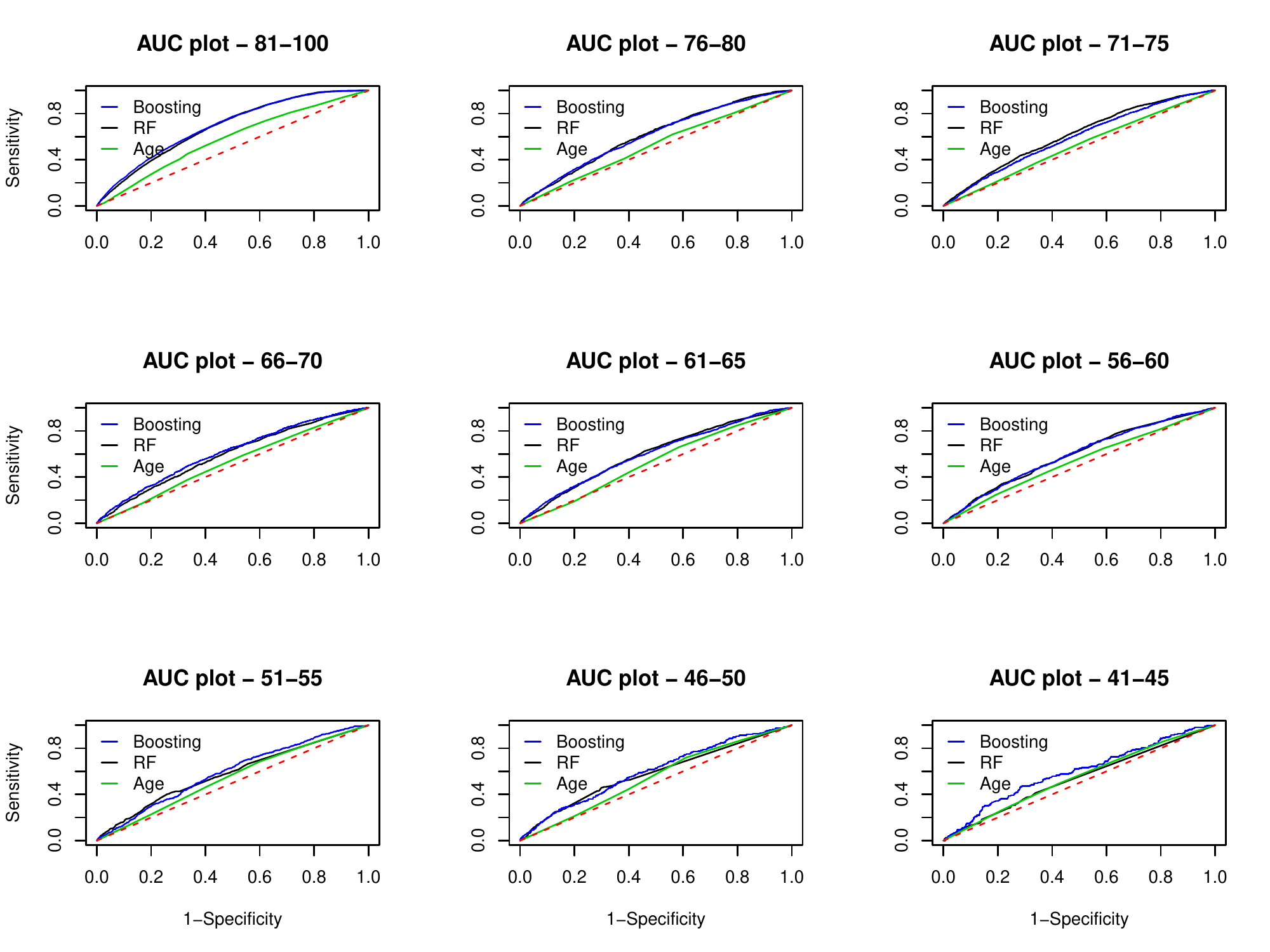}
\end{figure}

\begin{figure}[htb]
\caption{Out-of-Sample Area Under The Curve - 2015}
\label{f:auc2015}
\includegraphics[width=0.95\textwidth]{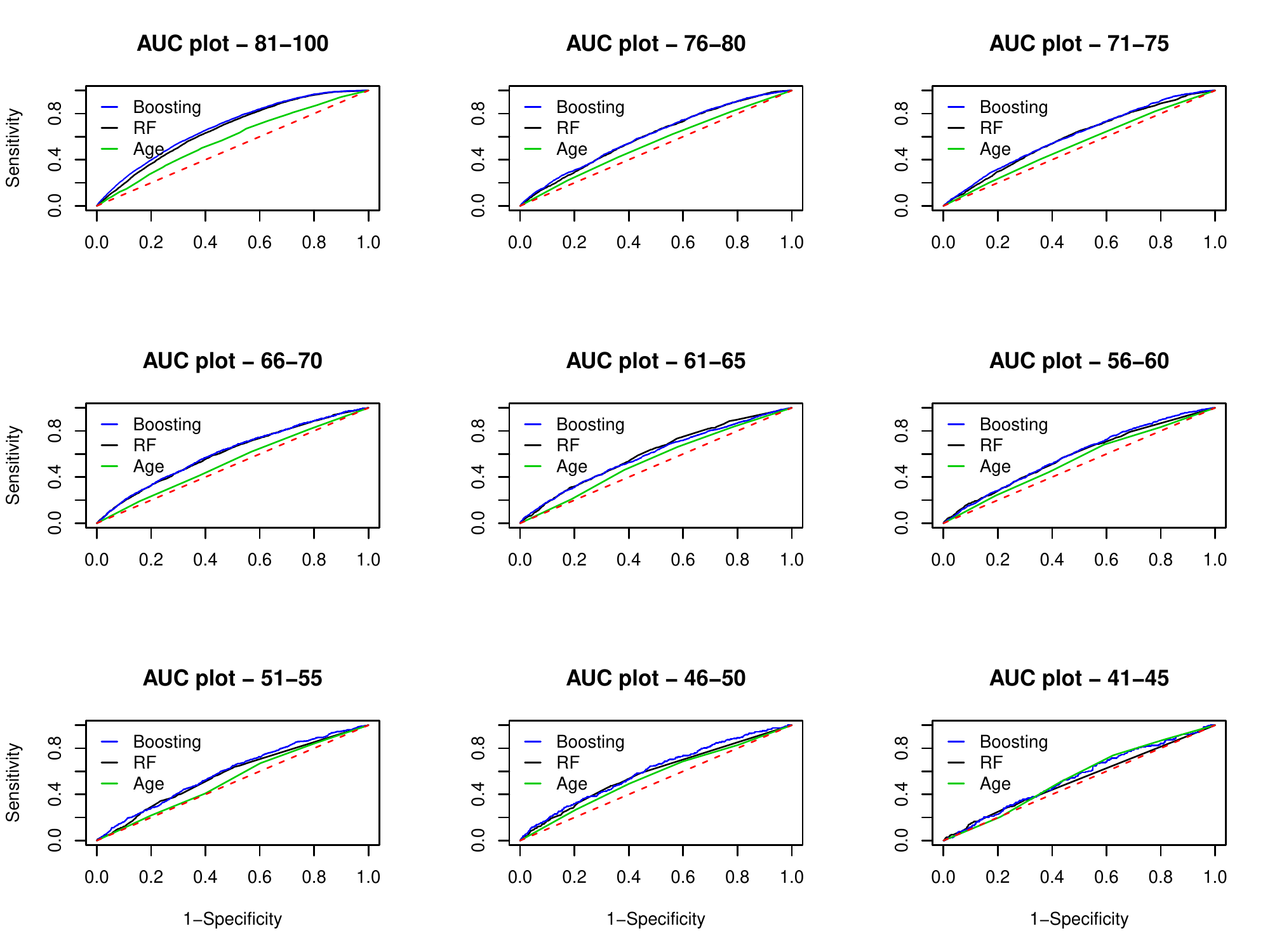}
\end{figure}

Figures (\ref{f:auc2013} - \ref{f:auc2015}) show the AUC figures for the years of 2013 to 2015 that were omitted from the main document. The results are very similar across all years. Random Forest and Gradient Boosting with credit variables have bigger Areas Under the Curve (AUC) than the age model, especially for cohorts of older individuals.

Table \ref{tab:sample} shows some information about the sample size in each out-of-sample year. Columns 2 and 3 show the number of deaths overall and the number of deaths of people older than 40 years old. Columns 4 and 5 show the sample size overall and the sample size for people older than 40 years old. 

\begin{table}[htb]
\caption{Sample size and number of deaths}
\label{tab:sample}
\begin{tabular}{lcccc}
\hline
year of death & N. Deaths & N. Deaths 40+ & Sample Size & Sample Size 40+ \\ \hline
2012          & 8,584     & 8,366         & 2,249,621   & 1,636,823       \\
2013          & 9,330     & 9,094         & 2,241,037   & 1,672,928       \\
2014          & 9,965     & 9,730         & 2,231,707   & 1,698,597       \\
2015          & 12,507    & 12,231        & 2,221,742   & 1,721,584       \\
2016          & 13,296    & 12,999        & 2,209,235   & 1,740,088       \\ \hline
\end{tabular}
\end{table}

\newpage

\newpage

\newpage

\end{document}